\documentclass[12pt]{article}
\usepackage{geometry}
\usepackage{graphicx} 
\usepackage[toc,title,page]{appendix}
\usepackage{cite} 
\usepackage{amsfonts}
\usepackage{amsmath}
\usepackage{amssymb}
\usepackage{physics}
\usepackage[dvipsnames]{xcolor}
\usepackage{feynmp-auto}
\usepackage{epsfig}
\usepackage{slashed}
\allowdisplaybreaks
\usepackage{comment}
\usepackage{authblk}
\usepackage{blindtext}
\usepackage{hyperref}
\usepackage{cleveref}
\usepackage{setspace}
\usepackage{tikz}
\usepackage{lipsum}
\usepackage{ragged2e}
\usepackage{xcolor}
\usepackage{soul}
\usepackage[c]{esvect}
\usepackage{harpoon}
\usepackage{subcaption}
\usetikzlibrary{angles,quotes}
\usepackage[english]{babel}
\usepackage[autostyle]{csquotes}
\usepackage{nicematrix}
\usepackage{blkarray}
\usepackage{float}

\newcommand{\bc}{\begin{center}}
\newcommand{\ec}{\end{center}}

\newcommand{\ra}{\rangle}
\newcommand{\la}{\langle}
\newcommand{\nn}{\nonumber}
\newcommand{\mc}{\mathcal{C}}
\newcommand{\md}{\mathcal{D}}
\newcommand{\mt}{\mathcal{T}}
\newcommand{\mm}{\mathcal{M}}
\newcommand{\p}{A B}
\newcommand{\pp}{C D}
\newcommand{\q}{B C}
\newcommand{\ca}{\cos \alpha}
\newcommand{\sa}{\sin \alpha}

\title{ Regularized Compton double scattering via unitarity}
\author{Shanmuka Shivashankara$^1$\footnote{sshivash@providence.edu} 
, Isra Gashi$^2$ \footnote{Corresponding author: isragashi@uni-pr.edu, } }
\affil{\itshape\small $^1$
Providence College, \itshape\small Providence, RI\ 02918  USA
}\affil{\itshape\small $^2$ University of Prishtina, \itshape\small Prishtinë 10 000  Republic of Kosovo
}

\date{
\small }

\numberwithin{equation}{section}

\begin{document}

\maketitle
\begin{abstract} When two initially entangled photons each undergo Compton scattering, the scattered electrons become correlated. However, the final reduced density matrix of one scattered pair is not influenced by the other scattered pair due to unitarity.  Herein, we keep unitarity up to tree level for Compton double  scattering and obtain different results than recent literature.  The initial four particles, where the initial photons are entangled, are written as a superposition of two states with a relative phase.  The final density matrix  has two area divergences that are regularized with unitarity.  The regularization procedure, i.e. solving for the roots of a polynomial that represents the probability for no scattering, suggests a novel definition of the scattering cross-section.  Vieta's formulas relate these divergences to finite cross-sections.  For an initial pure state, the formulas for the final density matrix and the correlation of final electronic polarizations are given.  The correlation implies double scattering is analogous to Young's diffraction experiment.  The two initial superposed states are the circular apertures while the Feynman amplitudes are the interfering complex light fields.
\end{abstract}



\section{Introduction}

In entanglement swapping, there exists entanglement within two separate pairs of particles, say $\p$ and $\pp$, but not between the pairs.  After particles $BC$ scatter, particles $AD$ are still separable because of unitarity.  See section 2, item ($iv$), in \cite{shiva4}.  However, a projection of the final state onto a Bell state allows particles $AD$ to be inseparable.  Here, we study a double scattering process that is the complement of entanglement swapping.  Unlike entanglement swapping, the initial entanglement only exists between two particles from two different pairs of particles that undergo scattering.  In this paper, the term $(Compton)\ double\ scattering$ refers to the complement of entanglement swapping.  However, when referring to a density matrix, the term double scattering refers to the following subspaces: no scattering, single scattering, and double scattering.  

Recently, \cite{Fan3} calculates the final density matrix for double scattering without keeping unitarity and regularizing divergences.  By keeping unitarity the final density matrix should include the following subspaces: no scattering, single scattering, and double scattering.  Unitarity also suggests the regularization of divergences in density matrices for decays or scattering processes \cite{shiva2,shiva3,shiva4}.  However, \cite{Fan3} omits single scattering and forward scattering terms, which would give the probability for no scattering.  Our work remedies that omission.  They also conflate particles across subspaces when calculating the   mutual information for double scattering.  \cite{shiva4} provides a brief summary of unitarity's importance in particle interactions.

In section \ref{two}, unitarity is shown to govern entanglement generation in (Compton) double scattering and requires different results than \cite{Fan3}. See \cite{shiva1} for unitarity's role in Compton scattering when the scattering photon is entangled with another photon (witness) that does not partake in an interaction.  Unitarity requires that the scattering leaves the witness photon's reduced density matrix unchanged.  However, the witness and scattered electron become correlated since the mutual information is nonzero.   

In section \ref{subspaces}, we keep unitarity up to tree level when calculating the final density matrix for Compton double scattering.   In section \ref{regular}, the final density matrix has two area divergences that are shown to be regularized with unitarity, i.e.\ finding the roots of the probability for no scattering.  Vieta's formulas relate these areas to the total and double cross-sections.  This novel regularization procedure suggests an alternative definition of the scattering cross-section.  For $n$ Compton scatterings of $n$ entangled photons with $n$ electrons, the roots can be imaginary.  However, Vieta's formulas relate these roots to a real total or $n$-tuple scattering cross-section.

Double scattering provides further confidence in the procedure for regulating divergences (cf. \cite{shiva2,shiva3}).  However, others do not regularize $\dfrac{\delta^3(0)}{\delta(0)}$ in terms of the scattering cross-section and assert it is indeterminate \cite{Fan3, kow1,kow2}.  See appendices A1 and A2 in \cite{shiva3} for an argument that the divergence is not indeterminate.  Also, for an inelastic scattering process, the regularization procedure implies that a particle's von Neumann entanglement entropy of momentum reduces to the Shannon entropy if the quantum mechanical contribution goes to zero (see section 4 in
\cite{shiva4}).


In section \ref{red}, the final density matrix of the two scattered electrons is presented for a specific initial spin configuration of the two electrons and two photons.  From this formula, the correlation between electronic polarizations is calculated and seen to evoke Young's diffraction experiment.  Interference among the Feynman amplitudes or initial states defines the $complex\ degree\ of\ coherence$, which is a well known parameter in interference experiments in optics.  The presentation is general enough to be adapted to other double scattering interactions.       

\section{Unitarity's constraint on double scattering}\label{two}

If four particles have no entanglement among them, the scattering in one pair should have no bearing on the other pair.  Unitarity confirms this intuition as follows.  Let the four particles have an initial state $|i\ra = |A\ra \otimes B\ra \otimes |C\ra \otimes D\ra$.  Each particle's normalized ket, e.g. $|A\ra$, contains the particle's momentum and angular momentum.  Suppose the two pairs, $AB$ and $CD$, undergo possible scattering, producing a final state $|f\ra = (S_1 |AB\ra) \otimes (S_2 |CD\ra)$.  $S_{1,2}$ are the unitary $S$-matrices for the two scatterings.  

The final density matrix is
\bc
$| f\ra \la f| = (S_1 |AB\ra \la AB|S^\dagger_1) \otimes (S_2 |CD\ra \la CD|S^\dagger_2)$.
\ec
Trace the above over all possible states for system $\pp$.  The remaining density matrix becomes the reduced density matrix for system $\p$, 
$\rho^f_{AB}= S_1 |AB\ra \la AB| S^\dagger_1$.  Therefore, any interaction in system $\pp$ has no influence on scattering in system $\p$.  This means unitarity, i.e. $S_2 S_2^\dagger=1$, requires the    von Neumann entanglement entropies, mutual information,  correlations within system $\p$ to be the same regardless of scattering in system $\pp$.  The entropy is given by $S_k=-Tr(\rho_k \log \rho_k)$, where $k$ is the system of interest, ~e.g. system $\p$.  The mutual information of $\p$'s polarizations is defined as $I(A,B)$ = $S_A + S_B -S_{AB}$ and must be zero when the system is separable.

Now consider the scenario of Compton double  scattering with initial spin entanglement in system $BC$, say a pair of photons.  Let particles $A$ and $D$ refer to electrons.  A general initial state is
\begin{align}\label{rhoi}
|i\ra = |A\ra \otimes \Big(\ca\ |B_1\ra \otimes | C_1\ra + e^{i\beta} \sa\ |B_2\ra \otimes |C_2\ra \Big) \otimes |D\ra.
\end{align}
$B_1B_2$ above refer to the same photon $B$, but of opposite helicities.  Similarly, $C_1C_2$ refer to the same photon $C$.  $\alpha$ and $e^{i\beta}$ are the entanglement parameter and relative phase, respectively.

For double scattering, apply the unitary operators $S_1S_2$ on eq. (\ref{rhoi}). Tracing all possible  states for system $\pp$, the final density matrix for system $\p$ after scattering is
\begin{align}\label{rhof12}
\rho^f_{AB} &= \cos^2 \alpha\ S_1| A;B_1\ra \la A;B_1|S_1^\dagger + \sin^2\alpha\ S_1| A; B_2\ra \la A; B_2|S_1^\dagger\\
&= S_1 \rho^i_{AB} S_1^\dagger,\nn
\end{align}
where $\rho^i_{AB}$ is the initial reduced density matrix for particles $AB$.  Equation (\ref{rhof12}) holds regardless of whether the initial state is pure or mixed.  Therefore, the final density matrix of system $\p$ is unaffected by the interaction in system $\pp$ in spite of initial entanglement between particles $\q$.  This is due to unitarity.  Likewise, the mutual information within the scattered system $\p$ is unaffected.  However, Ref. \cite{Fan3} finds otherwise due to not keeping unitarity.  They considered double $e^+e^-$ scattering where the initial $e^-$'s are entangled.  $S_1S_2$ above captures the following four subspaces: both particle pairs scatter, neither scatter, or only one pair scatters.  However, eq.~(7) in \cite{Fan3} only allows for both pairs scattering and no scattering.  Also, their mutual information of polarizations conflates particle identities across subspaces.     

When calculating the final density matrix, \cite{Fan3} omits two $types$ of forward scattering terms in their eq.~(7).  One type would allow the final density matrix to have a term  representing the probability for no scattering.  The other omission pertains to their integral over final states dropping the forward scattering.  The latter is understandable given a collider's lack of detection along the beam line. If only considering experimental trials with both particle pairs scattering, they should drop the initial state or first term in their eq.~(7).  In this limited subspace, scattering in system $AB$ is affected by scattering in system $CD$.  Otherwise, both forward scattering of the first type and single scattering events should be included.  See eq.~(\ref{rhof}) and its interpretation below.

\section{Double scattering via unitarity}\label{subspaces}

We now consider Compton double scattering. Scattering occurs within particle pairs $AB$ and $CD$ where only particles $BC$ (photons) have initial entanglement.  The divergences that occur in the final density matrix are regularized in section \ref{regular}.  The final density matrix and correlation of $AD$ (electrons) is given in section \ref{red}.  The unitary $S$-matrices are $S_j = 1+ i\mt_j,$ for $j=1,2$, where $\mt_j$ is the transition operator.  

The final density matrix for double scattering, $\rho^f$, can be grouped into a direct sum of subspaces as follows.
\begin{align}\label{rhof}
\rho^f &= S_1 S_2 \rho^i S_2^\dagger S_1^\dagger\nn \\
&\equiv N + S + \md,  
\end{align}

where
\begin{align}\label{N}
N =&\ \rho^i + (i\mt_1 \rho^i +H.c.) + (i\mt_2 \rho^i + H.c.)\\
&\ + \Big( i\mt_1\ ( i\mt_2 \rho^i - \rho^i i\mt_2^\dagger) + H.c. \Big),\nn
\end{align}
\begin{align}\label{S}
S =&\ (i\mt_1 \rho^i (i\mt_1 i\mt_2)^\dagger  + H.c.) + (i\mt_2 \rho^i (i\mt_1 i\mt_2)^\dagger  + H.c.) \\
&\ + (i\mt_1) \rho^i (i\mt_1)^\dagger + (i\mt_2) \rho^i (i\mt_2)^\dagger,\nn
\end{align}
\begin{align}\label{D}
\md =&\ i\mt_1 i\mt_2 \rho^i (i\mt_1 i\mt_2)^\dagger. 
\end{align}

  $N$ above refers to no scattering having occurred.  $S$ refers to single scattering, i.e.\ either $AB$ or $CD$ scattering, but not both having occurred.  $\md$ refers to double scattering, i.e.\ both pairs of particles undergoing scattering.  If the final density matrix was limited to $\md$ above, the reduced density matrix for final particles $AB$ would be dependent on the scattering of particles $CD$.  This result means the mutual information between scattered pairs $AB$ and $CD$ would be nonzero.  However, keeping unitarity (subspaces $N,S$) the scattering in system $CD$ has no bearing on the system $AB$ on $average$.  See eq.~ \ref{rhof12}.     

Analyzing the coherence generation within the final system $AD$ requires tracing the above final density matrix.  This gives rise to a divergent volume factor, $V = (2\pi)^3\delta^3(0)$, and a time divergent factor,  $T = 2\pi\delta(0)$, in the reduced density matrices.  The arguments in $\delta^3(0)$ and $\delta(0)$ have units of momentum and energy, respectively.  These divergences appear together as the ratio $\dfrac{V}{\upsilon T}$, which has the appearance of a scattering cross-section or interaction rate, $1/T$, divided by the luminosity $\upsilon/V$.  $\upsilon$ is the relative velocity between colliding particles.  The regularization follows from the probability of no scattering, i.e.\ the trace of $N$ in eq. (\ref{N}) over all initial particles  \cite{shiva3,shiva4}.  For double scattering, two regularizations occur as follows.  

\subsection{Regularization of double scattering}\label{regular}

Suppose the initial state of four particles, $ABCD$, is mixed.  Particles $BC$ are initially entangled via an entanglement parameter, $\alpha$.  A particle's ket, e.g. $|A\ra$, contains the particle's momentum and polarization information.  Then, the initial density matrix is given by
\begin{align}\label{arhoi}
\rho^i = |A;D\ra \otimes\Big(\cos^2 \alpha |B_1;C_1\ra \la B_1;C_1| + \sin^2 \alpha |B_2;C_2\ra \la B_2;C_2|\Big)\otimes \la A;D|.\end{align}
After possible scattering in systems $AB$ and $CD$, eqs.~(\ref{rhof})-(\ref{D}) give the final density matrix.  Subspace $N$ in eq.~(\ref{N}) represents the initial states.  Tracing over these initial states and using the optical theorem gives the probability for no scattering, 
\begin{align}\label{reg}
1-\Big(\dfrac{\sigma_{AB}}{\frac{V}{\upsilon T}} + \dfrac{\sigma_{CD}}{\frac{V}{\upsilon T}}- \dfrac{\sigma_{AB,CD}}{(\frac{V}{\upsilon T})^2} \Big),
\end{align}
where
\begin{align}\label{sigmamixed}
\sigma_{AB}\equiv&\  \cos^2\alpha \sum_f \sigma_{AB_1\rightarrow f} + \sin^2 \alpha \sum_f \sigma_{AB_2\rightarrow f},\nn\\
\sigma_{CD}\equiv&\ \cos^2 \alpha \sum_f \sigma_{C_1D\rightarrow f} + \sin^2 \alpha \sum_f \sigma_{C_2D\rightarrow f},\nn\\
\sigma_{AB,CD}\equiv&\  \cos^2\alpha \sum_f \sigma_{AB_1\rightarrow f}\sum_f \sigma_{C_1D\rightarrow f} +  \sin^2 \alpha \sum_f \sigma_{AB_2\rightarrow f}\sum_f \sigma_{C_2D\rightarrow f}.
\end{align}
$\sigma_{AB_i\rightarrow f}$ ($\sigma_{C_i D\rightarrow f}$) is the total cross-section of particles $AB_i$ ($C_iD$) scattering into final state $f$.

To obtain the regularization of $\dfrac{V}{\upsilon T}$, set the probability for no scattering, i.e. eq. (\ref{reg}), to an arbitrary constant $\xi$, such that $\xi>0$.  For the mixed $\rho^i$ in eq.(\ref{arhoi}), the regularizations are 
\begin{align}\label{regmixed}
\dfrac{V}{\upsilon T} = \dfrac{(\sigma_{AB}+\sigma_{CD}) \pm \sqrt{(\sigma_{AB}+\sigma_{CD})^2-4\sigma_{AB,CD}(1-\xi)}}{2(1-\xi)}.    
\end{align}
The discriminant above is positive if $\xi > 1-\dfrac{(\sigma_{AB} + \sigma_{CD})^2}{4\sigma_{AB,CD}}$, which implies $\xi$ must be less than one.  Adding the two regularizations in eq.~(\ref{regmixed}) gives $\dfrac{\sigma_{AB}+\sigma_{CD}}{(1-\xi)}$.  Also, the product of the two regularizations is $\dfrac{\sigma_{AB,CD}}{(1-\xi)}$.  These two relations are well known as Vieta's formulas, which relate the roots (regularizations) of a polynomial equation to the coefficients.   

  In general the divergences $\dfrac{V}{\upsilon T}$ stemming from an arbitrary final density matrix or scattering process should carry meaning. For example, regularizations were recently obtained for a weak decay and a scattering process with a witness.  Those regularizations correctly refer to the total decay width \cite{shiva2} and the scattering cross-section \cite{shiva3}.  Also, in \cite{shiva4}, the regularization procedure gives the correct cross-section for a spinless particle scattering from an impenetrable sphere without using a scattering amplitude or Lippmann-Schwinger equation plus Green's function.    
  
  As another example, consider eq. (\ref{arhoi}). If $B_1 = B_2$ and $C_1 = C_2$ , then eq.\ (\ref{regmixed}) implies the two regularizations would be $\sigma_{AB}$ and $\sigma_{CD}$ with $\xi=0$.  Since the initial states $AB$ and $CD$ are separable, the total and double cross-sections or probabilities would be the sum and product of the roots, i.e.\ $\sigma_{AB}+ \sigma_{CD}$ and $\sigma_{AB} \sigma_{CD}$, respectively. This suggests that Vieta's roots for eq.\ (\ref{arhoi}) pertain to independent areas.  Recall that unitarity requires the system $AB$ to be independent of $CD$ (see section 2 above).  This requires that Vieta's sum of roots gives the total probability, $\dfrac{\sigma_{AB}+\sigma_{CD}}{(1-\xi)}$, where $\xi=0$ for a physical cross-section.  As $\xi\rightarrow 1$, the total cross-section diverges and may indicate absorption.  The relevance of the individual roots by themselves is not known.  The regularization, i.e.\ solving for the roots of the probability for no scattering, may be considered an alternative definition for scattering cross-sections.    
  

 Double scattering can be generalized to $n$-pairs of Compton scatterers where the $n$ photons are initially entangled.  The final subspaces will be no scattering, single scattering, double scattering, ..., $n$ scattering. e.g., let $n=3$ and label the six particles as $ABCDEF$.  Set the probability for no scattering to $\xi$ and define the roots as $r\equiv \dfrac{V}{\upsilon T}$.  The three roots, $r_i,$ $i=1,2,3$ are related to the cross-sections as follows.
\begin{align*}
 r_1 + r_2 + r_3  =& (\sigma_{AB} + \sigma_{CD} + \sigma_{EF})/(1-\xi),\\
 r_1 r_2 + r_1 r_3 + r_2 r_3 =&  (\sigma_{AB,CD} + \sigma_{AB,EF} + \sigma_{CD,EF})/(1-\xi),\\
 r_1 r_2 r_3 =& \sigma_{AB,CD,EF}/(1-\xi).
 \end{align*}
 These are Vieta's formulas for a cubic polynomial,  $r^3 - b\ r^2 + c\ r -d = 0$, where $b = (\sigma_{AB} + \sigma_{CD} + \sigma_{EF})/(1-\xi)$, $c = (\sigma_{AB,CD} + \sigma_{AB,EF} + \sigma_{CD,EF})/(1-\xi)$, $d = \sigma_{AB,CD,EF}/(1-\xi)$.  The first root is 
 \begin{align*}
 r_1 =&\ \frac{b}{3} + \frac{\dfrac{b}{3} - \dfrac{c}{b}}{\Big(1 - \frac{9c}{2b^2} + \frac{27d}{2b^3} + \frac{3\sqrt{3}}{2} \sqrt{- \frac{c^2}{b^4} + 4\frac{c^3}{b^6} + 4\frac{d}{b^3} - 18\frac{cd}{b^5} + 27\frac{d^2}{b^6}}\ \Big)^{1/3}}\\
 & + \frac{b}{3}\Big(1 - \frac{9c}{2b^2} + \frac{27d}{2b^3} + \frac{3\sqrt{3}}{2} \sqrt{- \frac{c^2}{b^4} + 4\frac{c^3}{b^6} + 4\frac{d}{b^3} - 18\frac{cd}{b^5} + 27\frac{d^2}{b^6}}\ \Big)^{1/3}\\
 =& b - \frac{c}{b} + ...\ .
 \end{align*}
The remaining two roots $r_2,r_3$ are the complex conjugate of one another.  Adding these  two roots gives $r_2+r_3=c/b +...$ whereas the magnitude of the difference is $|r_2-r_3|=\dfrac{3\sqrt{3} d}{2b^2}+...$.  For $n$ being odd, there is always at least one real root.  
 


For an initial pure state (eq.\ (\ref{rhoi})) and following the same procedure above, the regularization
$\Big( \dfrac{V}{\upsilon T}\Big)_{\text{pure}}$
 has the same form as eq.~(\ref{regmixed}).  However, $\sigma_{AB,CD}$ (see eq.\ (\ref{sigmamixed})) would be missing interference between initial states $AB_1$ ($C_1 D$) and $AB_2$ ($C_2 D$).  With $\mathcal{M}(\cdot)$ understood as the Feynman amplitude and $d\Pi_f$ denoting the phase space integral \cite{peskin}, 
 \begin{align}\label{sigmapure}
 (\sigma_{AB,CD})_{pure} =& \sigma_{AB,CD} + \frac{\sin 2\alpha}{2}(\frac{e^{-i\beta}}{2E_A E_B \upsilon} \sum_f \int d\Pi_f\ \mathcal{M}(AB_2\rightarrow f)^\dagger \mathcal{M}(AB_1\rightarrow f)*\nn\\
 &\quad \frac{1}{2E_C E_D \upsilon} \sum_{f'} \int d\Pi_{f'}\ \mathcal{M}(C_2D\rightarrow f')^\dagger \mathcal{M}(C_1D\rightarrow f')  + H.c.)
 \end{align}

\subsection{Final density matrix and correlation of electronic polarizations in Compton double scattering}\label{red}

\begin{figure}[H]
    \centering
    \includegraphics[width=.5\textwidth]{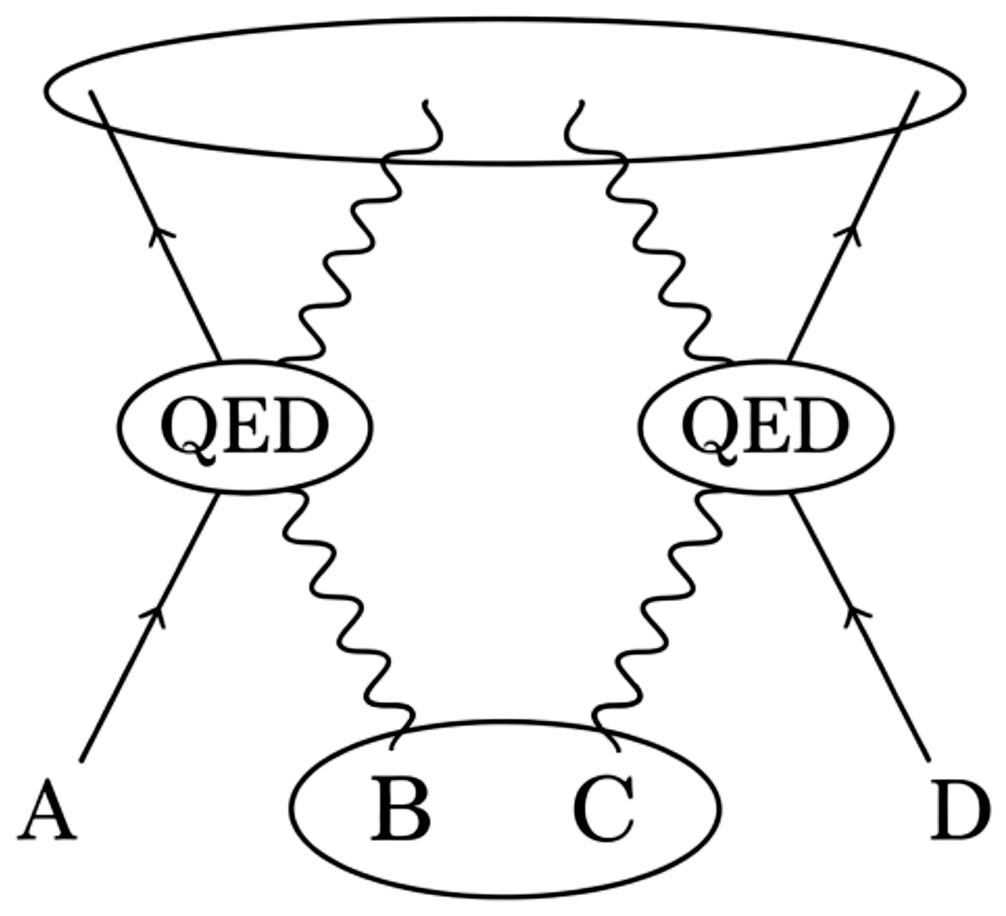} 
    \caption{Initially particles BC are entangled.  After the initial particle pairs AB and CD scatter, coherence is generated in the final system $AD$.}
    \label{fig1}
\end{figure}

Scattering spreads coherence from the initial entangled particles $BC$ to the final particles $AD$.  See Figure \ref{fig1}.  Notice there are two final particles of type $A$ ($D$) since there are multiple interaction subspaces, namely $S$ and $D$.  Due to tracing over all initial particles and final particles $BC$ and using the regularization, all subspaces in eq. (\ref{rhof}) are relevant.  The matrix is a direct sum of four terms.  $\xi$, the probability for no scattering, is the first term.  The second (third) term refers to final state particle $A$ ($D$) after single scattering.  The fourth term refers to final state particles $AD$ from double scattering.  Only the latter or subspace $\md$ is relevant for obtaining spin correlations or the mutual information between final particles $AD$.  If the $S$ subspace was added, the spin correlation would remain unchanged while the mutual information would conflate particle identities in the different subspaces $S$ and $D$.  

Suppose two pairs of particles undergo Compton scattering.  The initial four particles have momentum-polarization pairs $(\vb*p_i, r_i)$, $i=1,2,3,4$.  The final  particles have momentum-polarization pairs $(\vb*q_i,s_i)$, $i=1,2,3,4$.    Let the initial pure state be written with the initial momenta, $\vb*p_i$, suppressed.
\begin{align}\label{rhoir}
|i\ra = |r_1\ra \otimes \Big(\ca\ |r_2\ra \otimes | r_3\ra + e^{i\beta} \sa\ |r_3\ra \otimes |r_2\ra \Big) \otimes |r_4\ra    
\end{align}
In both terms above, the four kets are ordered to correspond to the four particles $ABCD$.  Particle $A$ ($D$) has the initial polarization $r_1$ ($r_4$).  Particles $BC$ are entangled.  The entangled particle $B$ ($C$) has either the initial polarization $r_2$ ($r_3$) or $r_3$ ($r_2$).  The initial electrons ($AD$) are at rest.  The initial photons both have energy $\omega$ while the final photons have energies $\omega_i,\ i=2,3$.

The final reduced density matrix of electronic polarizations for system $AD$ is taken from subspace $\md$ in eq. \ref{D} and written as 
\begin{align}\label{rho14}
\rho^f_{AD}=&\  \frac{1}{N} \sum_{s_1,s^{'}_1,s_4,s^{'}_4} |s_1,s_4\ra \la s'_1,s'_4|*\nn\\ 
& \sum_{s_2,s_3}\Big(\cos^2 \alpha\  \mc_{r_1,r_2;r_1,r_2}^{s_1,s_2;s'_1,s_2}\ \tilde{\mc}_{r_3,r_4;r_3,r_4}^{s_3,s_4;s_3,s'_4} +   \sin^2\alpha\ \mc_{r_1,r_3;r_1,r_3}^{s_1,s_2;s'_1,s_2}\ \tilde{\mc}_{r_2,r_4;r_2,r_4}^{s_3,s_4;s_3,s'_4}\nn \\
 &\ + \frac{\sin2\alpha}{2}\ e^{-i\beta}\ \mc_{r_1,r_2;r_1,r_3}^{s_1,s_2;s'_1,s_2}\ \tilde{\mc}_{r_3,r_4;r_2,r_4}^{s_3,s_4;s_3,s'_4} + \frac{\sin2\alpha}{2}\ e^{i\beta}\ \mc_{r_1,r_3;r_1,r_2}^{s_1,s_2;s'_1,s_2}\ \tilde{\mc}_{r_2,r_4;r_3,r_4}^{s_3,s_4;s_3,s'_4} \Big),  
\end{align}
 where $N=Tr(\rho^f_{AD})=(\sigma_{AB,CD})_{pure}$ (see eq.\ (\ref{sigmapure}) is the normalization constant and
 \begin{align}\label{cs}
 \mc_{r_1,a;r_1,b}^{s_1,s_2;s'_1,s_2} \equiv&  \dfrac{1}{2E_A E_B \upsilon} \int d\Pi_2\ \mm^{s_1s_2}_{r_1a}  (\mm^{s'_1s_2}_{r_1b})^\dagger,\nn\\      \tilde{\mc}_{a,r_4;b,r_4}^{s_3,s_4;s_3,s'_4} \equiv&  \dfrac{1}{2E_C E_D \upsilon} \int d\Pi_2\  \mm^{s_3s_4}_{a\ r_4}  (\mm^{s_3s'_4}_{b\ r_4})^\dagger.
 \end{align}
Notice $\mc_{r_1,a;r_1,b}^{s_1,s_2;s'_1,s_2}$ gives the polarized Thomson scattering cross-section ($\omega\rightarrow 0$) when $a=b$ and $s_1=s'_1$ \cite{peskin}.  $\rho^f_{AD}$ above is a linear sum of polarized double cross-sections.  The general eq.~(\ref{rho14}) allows for any type of interactions within particle pairs $AB$ and $CD$ given initial entanglement between particles $BC$.  

\begin{figure}[H]
    \centering
    \includegraphics[width=1.0\textwidth]{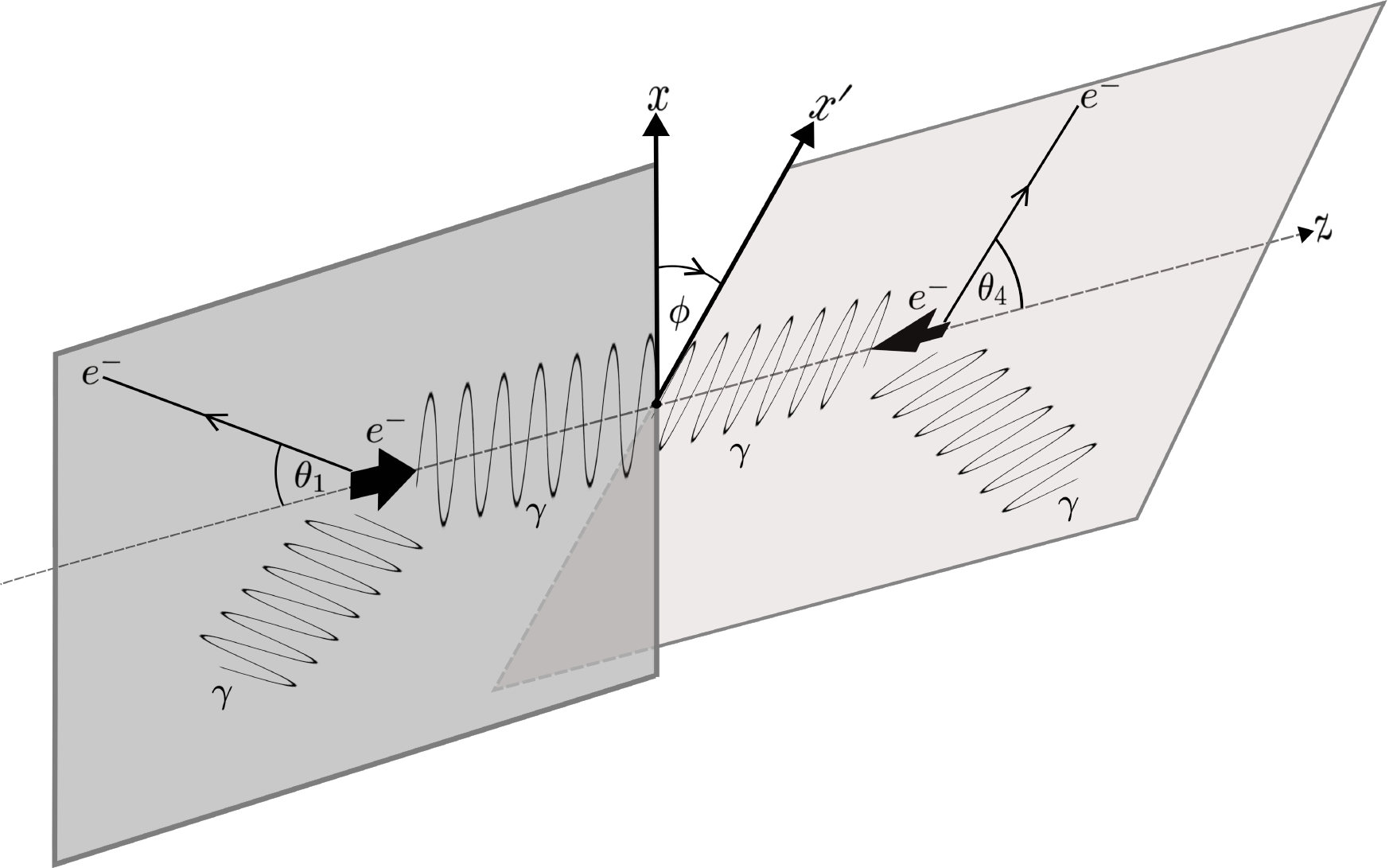} 
    \caption{Two entangled photons, $\gamma$, travel in opposite directions  from the origin and scatter from at-rest polarized electrons, $e^-$.  The large bold arrows along the $z$-axis indicate the electronic polarizations.  $\theta_1$ and $\theta_4$ are the scattering angles of the electrons relative to the negative and positive $z$-axis, respectively.  The scattered $e^- \gamma$ pair on the far left (right) spans the $xz$ plane ($x'z$ plane).  The $x'z$ plane makes an angle $\phi$ with respect to the $xz$ plane.}
    \label{fig2}
\end{figure}

$\mm^{cd}_{ab}$ above is the Feynman amplitude for Compton scattering.  For system $AB$,  
\begin{align}\label{su}
i \mm^{s_1 s_2}_{r_1 b}\ \  = \ \ &\bar{u}_{s_1}(\vb*q_1)
   (-ie \slashed{\varepsilon}_{s_2}^* (\vb*q_2)
   )\ i\frac{(\slashed{p}_1 +\slashed{p}_2+m)}{2 p_1 \cdot p_2}\
   (-ie \slashed{\varepsilon}_b (\vb*p_2))
   u_{r_1}(\vb*p_1)\\
  \nonumber\\
   \ +\ & \bar{u}_{s_1}(\vb*q_1)
   (-ie \slashed{\varepsilon}_b (\vb*p_2)
   )\ i\frac{(\slashed{p}_1 -\slashed{q}_2+m)}{-2 p_1 \cdot q_2}\
   (-ie \slashed{\varepsilon}_{s_2}^* (\vb*q_2))
   u_{r_1}(\vb*p_1)\nonumber,
\end{align}
where $u_{s_1}(\vb*q_1)=\dfrac{1+\gamma_5 \slashed{n}_{fA}}{2}u(\vb*q_1)$ and $u_{r_1}(\vb*p_1)=\dfrac{1+\gamma_5 \slashed{n}_{iA}}{2}u(\vb*p_1)$.  
Consider the case of initial electronic spins facing one another.  See Figure \ref{fig2}.  In the lab frame, the initial and final electronic spin quantization axes are $n^\mu_{iA}=(0,0,0,1)$ and $n^\mu_{fA}=(|\vb*q_1|/m,q_{10}/m\sin \theta_1,0,-q_{10}/m \cos \theta_1)$, respectively.  $m$ is the electron's mass.  In the scattered $e^-$'s rest frame, $n^\mu_{fA}\rightarrow (0,0,0,1)$.   In all inertial frames, $n_{fA}\cdot q_1=n_{iA}\cdot p_1=0$ and $n_{fA}^2 = n_{iA}^2 = -1$. The initial photon polarization vectors are $\epsilon_{b}^\mu(\vb*p_2) = \frac{1}{\sqrt{2}}(0,-1,\pm i,0)$ where the upper and lower signs refer to right-handed and left-handed helicities.  The scattered photon's polarization vectors are not specified since $\sum_{s_2=\pm} \epsilon_{s_2}^\mu \epsilon_{s_2}^{*\nu} \rightarrow -g^{\mu \nu}$.  For system $CD$, $u_{r_4}(\vb*p_4)=\dfrac{1+\gamma_5 \slashed{n}_{iD}}{2}u(\vb*p_4)$, $n^\mu_{iD}=-n^\mu_{iA}$, $n^\mu_{fD}=(|\vb*q_4|/m,q_{40}/m \cos\phi \sin\theta_4,q_{40}/m\sin\phi \sin\theta_4,q_{40}/m \cos \theta_4)$, and $\epsilon_{b}^\mu(\vb*p_3) = \frac{1}{\sqrt{2}}(0,1,\pm i,0)$. $\phi$ is the angle between the two scattering planes.  

eq. (\ref{rho14}) implies that the reduced density matrix of final particle $A$ is influenced by scattering in system $CD$. Compare this result with eq. (\ref{rhof12}) where the opposite holds.  The discrepancy lies in that eq.\ (\ref{rho14}) only considers double scattering, and not single scattering events.  The latter equation implies the final electronic polarizations have a mutual information that is nonzero.  This would mean the initial coherence within $BC$ can spread to $AD$  because of double scattering.

The correlation between final particles $AD$ is defined as $\la \sigma^z_A\otimes \sigma^z_D\ra - \la \sigma_A^z\ra \la \sigma_D^z\ra=Tr(\sigma^z_A\otimes \sigma^z_D\ \rho^f_{AD}) - Tr(\sigma^z_A\ \rho^f_{A})Tr(\sigma^z_D\ \rho^f_{D})$, where  $\sigma^z_{A,D}$ are the same third Pauli spin matrix $((1,0),(0,-1))$.  Since the tree level Feynman amplitudes for Compton scattering are real and only the diagonal elements of $\rho^f_{AD}$ contribute to the above correlation, the correlation is the ratio of linear functions of $\cos\beta$.  The first term is   
\begin{align}\label{interf}
\la \sigma^z_A\otimes \sigma^z_D\ra =&\ \dfrac{1}{N}  \sum_{s_1,s_4=\pm1/2}\ (-1)^{s_1-s_4}*\nn\\
&\quad \sum_{s_2,s_3} \Big(\cos^2\alpha\ \mc_{r_1,r_2;r_1,r_2}^{s_1,s_2;s_1,s_2}\ \tilde{\mc}_{r_3,r_4;r_3,r_4}^{s_3,s_4;s_3,s_4}+\sin^2\alpha\ \mc_{r_1,r_3;r_1,r_3}^{s_1,s_2;s_1,s_2}\ \tilde{\mc}_{r_2,r_4;r_2,r_4}^{s_3,s_4;s_3,s_4}\nn\\
&\ \  + \sin2\alpha\ \cos\beta\ \mc_{r_1,r_2;r_1,r_3}^{s_1,s_2;s_1,s_2}\ \tilde{\mc}_{r_3,r_4;r_2,r_4}^{s_3,s_4;s_3,s_4}\Big)\nn\\
\equiv &\ \dfrac{1}{N} \sum_{s_1,s_4=\pm1/2}\ (-1)^{s_1-s_4}*\nn\\
&\quad \sum_{s_2,s_3=\pm 1}\Big( I_1 + I_2 + 2\sqrt{I_1 I_2}\ \text{Re}(\gamma_{12})\Big),
\end{align}
 where the normalization, $N$, is given by eq.\ (\ref{sigmapure}). The double scatterings, $I_{1,2}$, and $complex\ degree\ of\ coherence$, $\gamma_{12}$, are defined as follows. 
\begin{align}\label{beta}
I_1 =&\ (\cos\alpha\  \mc_{r_1,r_2;r_1,r_2}^{s_1,s_2;s_1,s_2})\ (\cos\alpha\ \tilde{\mc}_{r_3,r_4;r_3,r_4}^{s_3,s_4;s_3,s_4})\nn\\
I_2 =&\ (\sin\alpha\ \mc_{r_1,r_3;r_1,r_3}^{s_1,s_2;s_1,s_2})\ (\sin\alpha\ \tilde{\mc}_{r_2,r_4;r_2,r_4}^{s_3,s_4;s_3,s_4}) \nn\\
\gamma_{12} =&\ \dfrac{\sqrt{\cos\alpha\sin\alpha}\ \mc_{r_1,r_2;r_1,r_3}^{s_1,s_2;s_1,s_2}\  \sqrt{\cos\alpha\sin\alpha}\ \tilde{\mc}_{r_3,r_4;r_2,r_4}^{s_3,s_4;s_3,s_4} }{\sqrt{I_1 I_2}}\ e^{i\beta}
\end{align}
The three terms in the parentheses of eq. (\ref{interf}) are the net double scatterings for a particular spin configuration of the final $e^-$'s and photons.  In optics, they are similar to the net irradiance from the interference between two complex light fields ($\vb*E_1,\vb*E_2$) that emanate from two circular apertures in an opaque screen \cite{hecht}.  Since an optic field is never monochromatic, $0 < |\gamma_{12}|< 1$ in free space.  The relative phase $e^{i \beta}$ in eq. (\ref{beta}) would be introduced by the circular apertures.  For us, the interference is between Feynman amplitudes $\mm^{s_1s_2}_{r_1r_2} \mm^{s_3s_4}_{r_3r_4}$ and $\mm^{s_1s_2}_{r_1r_3} \mm^{s_3s_4}_{r_2r_4}$ emanating from two initial superposed states with a relative phase.  See eq. (\ref{rhoir}).  Ostensibly, $0 < |\gamma_{12}|< 1$ due to spin uncertainty.  Also, only when $\beta=\pi/2$, are the double scatterings additive in eq. (\ref{interf}).  Therefore, Young's famous optic experiment and double scattering are analogous.  

If interested in the mixed state, or eq. (\ref{arhoi}), drop the last two terms in eq. (\ref{rho14}), which has the relative phase $e^{\pm i\beta}$.  When $\beta=\pi/2$ or the relative phase in eq. (\ref{rhoir}) is purely imaginary, an initial pure state gives the same interference between final particles $AD$ as does the mixed state.

\section{Discussion}

We studied coherence generation in Compton double scattering while keeping unitarity up to tree level.  Unitarity implies a larger final density matrix than the recent literature \cite{Fan3} and a means for regularization of divergences.  The initial two photons are entangled and scatter from different electrons.  Keeping unitarity requires that the total space of events includes no scattering, single scattering, and double scattering.  The polynomial or probability for no scattering has two area divergences, $\dfrac{V}{\upsilon T}$.  Regularizing the divergences, i.e.\ solving for the polynomial's roots, imply the roots are related to finite scattering cross-sections via Vieta's formulas.  The sum and product of roots give the probabilities for total and double scattering, respectively.  Ostensibly, the two regularizations are mutually exclusive areas.  The physical nature of these roots may warrant further study.  This procedure may be extended to having $n$ initial entangled photons partaking in Compton scatterings.  The sum and product of the $n$ roots or regularizations represent the probabilities for total and $n-tuple$ scattering, respectively.  This work is a fourth example of using this regularization procedure \cite{shiva2, shiva3, shiva4}.  Although divergences are a common occurrence when calculating final density matrices, the expectation values must be finite, necessitating a regularization procedure.   

Since the initial photons are entangled via a relative phase, the scattered electrons have coherence.  In section \ref{red}, the coherence of final electronic polarizations is due to the relative phase's real part or $\cos\beta$.  The $\cos\beta$ term reflects interference in the scattering channels or Feynman amplitudes.  After adjusting for bosonic versus fermionic spins, the general interference formula given in eqs.~(\ref{interf}, \ref{sigmapure}, \ref{cs}) may be applied to a variety of double scatterings.  Double scattering is analogous to Young's diffraction experiment.  The interference of two Feynman amplitudes via two initial superposed states corresponds to two complex light fields interfering via two circular apertures.

\setcounter{secnumdepth}{0}

\section{Acknowledgements}

Google Gemini (3.1 Pro) and OpenAI ChatGPT (5.5) provided useful discussions.

\section{Competing Interests}
Competing interests: The authors declare there are no competing interests.

\end{document}